%% file: main.tex
\documentclass{IEEEcsmag}

\usepackage[colorlinks,urlcolor=blue,linkcolor=blue,citecolor=blue]{hyperref}
\expandafter\def\expandafter\UrlBreaks\expandafter{\UrlBreaks\do\/\do\*\do\-\do\~\do\'\do\"\do\-}
\usepackage{upmath,color}
\usepackage{nameref}
\usepackage{flushend}

\jvol{XX}
\jnum{XX}
\paper{8}
\jmonth{Month}
\jname{IEEE Computing in Science and Engineering}
\jtitle{Publication Title}
\pubyear{2021}

\usepackage{url}
\usepackage{graphicx}%
\usepackage{multirow}%
\usepackage{amsmath,amssymb,amsfonts}%
\usepackage{amsthm}%
\usepackage{mathrsfs}%
\usepackage{xcolor}%
\usepackage{textcomp}%
\usepackage{manyfoot}%
\usepackage{booktabs}%
\usepackage{algorithm}%
\usepackage{algorithmicx}%
\usepackage{algpseudocode}%
\usepackage{tcolorbox}%
\usepackage{listings}%
\usepackage{lmodern}
\usepackage[normalem]{ulem}

\usepackage{subcaption}

\newcommand{\update}[1]{\textcolor{black}{#1}}

\begin{document}

%\title{Ten Simple Rules for Producing High-Quality Research Software}
\title{Ten Essential Guidelines for Building High-Quality Research Software}

\author{Nasir U. Eisty}
\affil{University of Tennessee, Knoxville, TN, 37916, USA}

\author{David E.~Bernholdt}
\affil{Oak Ridge National Laboratory, Oak Ridge, TN, 37831, USA}
% ORCID 0000-0001-7874-3064

\author{Alex Koufos}
\affil{Stanford University, Stanford, CA 94305, USA}

\author{David J. Luet}
\affil{Princeton University, Princeton, NJ, 08534, USA}

\author{Miranda Mundt}
\affil{Sandia National Laboratories, Albuquerque, NM, 87123, USA}

\begin{abstract}
High-quality research software is a cornerstone of modern scientific progress, enabling researchers to analyze complex data, simulate phenomena, and share reproducible results. However, creating such software requires adherence to best practices that ensure robustness, usability, and sustainability. This paper presents ten guidelines for producing high-quality research software, covering every stage of the development lifecycle. These guidelines emphasize the importance of planning, writing clean and readable code, using version control, and implementing thorough testing strategies. Additionally, they address key principles such as modular design, reproducibility, performance optimization, and long-term maintenance. The paper also highlights the role of documentation and community engagement in enhancing software usability and impact. By following these guidelines, researchers can create software that advances their scientific objectives and contributes to a broader ecosystem of reliable and reusable research tools. This work serves as a practical resource for researchers and developers aiming to elevate the quality and impact of their research software. 
\end{abstract}

\maketitle

%%%%% This notice is required by ORNL.  It needs to appear on the first page of the submitted manuscript.  
%%%%% It is okay for the publisher to remove it for publication.
\begin{figure}[!b]
\noindent\fbox{%
    \parbox{0.95\linewidth}{%}
This manuscript has been authored in part by UT-Battelle, LLC, under contract DE-AC05-00OR22725 with the US Department of Energy (DOE). The US government retains and the publisher, by accepting the work for publication, acknowledges that the US government retains a non-exclusive, paid-up, irrevocable, world-wide license to publish or reproduce the submitted manuscript version of this work, or allow others to do so, for US government purposes. DOE will provide public access to these results of federally sponsored research in accordance with the DOE Public Access Plan (\url{https://energy.gov/doe-public-access-plan}).
}
}
\end{figure}

\section{Introduction}
\label{sec_introduction}
\input{sec_introduction}

\section{Ten Essential Guidelines}
\label{sec_rules}
\input{sec_rules}

\section{Conclusion}
\label{sec_conclusion}
\input{sec_conclusion}

\section{ACKNOWLEDGMENTS}
\label{acknowledgments}
\input{acknowledgments}

\bibliographystyle{plain}
\bibliography{sn-bibliography}

\section*{Author Biographies}
\input{author-bios}

\end{document}

%% file: sec_introduction.tex
Developing high-quality research software has become a fundamental aspect of modern scientific practice. From data analysis and simulation to visualization and reproducibility, software is integral to advancing nearly every scientific discipline~\cite{Eisty2022}.
However, creating robust, reliable, and maintainable software poses significant challenges, particularly in research contexts where deadlines are tight, resources are limited, and priorities are often focused on novel results rather than sustainable tools~\cite{hannay2009howscientists}.
Consequently, a significant amount of research software suffers from issues such as lack of usability, poor documentation, difficulty in reproducibility, and inadequate long-term support.
These challenges highlight the critical need for clear guidelines and best practices to guide researchers in developing effective software~\cite{reproducibility_in_science,wilson2014bestpractices,wilson2017goodenough}.

Unlike commercial software, research software is often developed by scientists with limited formal training in software engineering~\cite{HEATON2015207,reproducibility_in_science}.
The primary goal of research software is to facilitate scientific discovery rather than achieve market readiness, which can result in compromises on code quality, maintainability, and performance~\cite{prabhu2011survey}.
Moreover, research projects often demand highly specialized software, making it difficult to rely on off-the-shelf solutions or existing tools~\cite{Eisty2022}.
These factors underscore the importance of equipping researchers with practical frameworks for producing software that meets the unique demands of scientific inquiry while adhering to quality and sustainability standards.

This paper addresses these challenges by presenting ten comprehensive guidelines for producing high-quality research software. 
The need for such guidelines has never been more urgent.
As the complexity and scale of research software increase, so does the reliance on software to handle large datasets, simulate intricate systems, and support reproducible workflows.
Poor-quality software hinders individual research efforts and undermines the broader scientific enterprise by making it difficult for others to reproduce and build upon results~\cite{peng2011reproducible}.
Conversely, high-quality research software enhances collaboration, accelerates discovery, and amplifies the reach of scientific findings.

\update{Research Software Engineers (RSEs) can play a crucial role in ensuring adherence to high-quality software development guidelines within research projects. 
Positioned at the intersection of scientific inquiry and software engineering, RSEs bring the technical expertise and methodological rigor needed to implement best practices that many domain scientists may lack the time or training to apply. 
Their involvement can help embed principles such as version control, modular design, continuous integration, and comprehensive testing into the development lifecycle practices often outlined in quality-focused guidelines but underutilized in typical research workflows. 
Moreover, RSEs can act as facilitators, translating high-level values into actionable strategies tailored to the specific needs of a project, and mentoring research teams in sustainable software development. 
By doing so, they not only enhance the reliability and reproducibility of individual projects but also foster a broader culture of software excellence in the research community.}

This paper aims to serve as a practical resource for researchers and developers who wish to improve the quality of their research software.
It draws on principles from software engineering, research methodology, and community best practices and adapts them to the specific needs of scientific research.
\update{While there have been previous papers of this kind (\cite{wilson2014bestpractices,wilson2017goodenough}), this paper is intended to present core values that can be translated as high-level practices, as opposed to finer delineated lists of individual actions.}
By following the ten guidelines outlined here, researchers can create software that is not only functional but also maintainable, extensible, and impactful.
In doing so, they contribute not only to their own projects but also to the advancement of research as a whole.
\update{The base guidelines are written for newly formed projects, but each section contains an example of how they may be modified for legacy projects.}

% \update{new text}

% \strike{deleted text}

%% file: sec_rules.tex
\subsection{\textbf{\textit{Guideline 1: Plan Before You Code}}}
\label{rule-plan}

Planning is the foundation of any successful software development project, particularly in research, where objectives can be complex and timelines tight.
Begin with a clear problem statement: What research problem does the software address?
What are its expected outputs, and how will users interact with it?
Define the requirements or scope explicitly to prevent scope creep, which can derail both timelines and focus.
This includes identifying the essential features of the initial version and distinguishing them from ``nice-to-have'' enhancements.

\update{To support the planning process, you may find it helpful to use Agile planning techniques such as creating user stories, defining sprints, and setting up a product backlog. Agile techniques can help manage the complexity and evolving nature of research software projects. Use project management tools like Trello or Asana to organize tasks and milestones, and consider using features like Kanban boards and Gantt charts for better visualization of project progress. This iterative approach allows for continuous refinement of project goals and adapts to changing requirements, ensuring that the software remains aligned with research objectives. \footnote{Want to learn more about Agile? Visit \url{https://www.atlassian.com/agile}}}

Developing a design or software architecture at this stage ensures a solid foundation.
\update{Software architecture refers to the fundamental structures of a software system and the discipline of creating such structures and systems.}
For instance, outline how different components of the software will interact and what data structures will be used.
Creating mockups for user interfaces or pseudocode for critical algorithms can further clarify the design.\footnote{Not sure how to get started? Visit \url{https://intersect-training.org/software-design/} for a good beginner's course.}
Additionally, anticipate challenges: What are the computational limits?
Are there ethical or legal concerns with the data?
Good planning includes strategies to mitigate potential risks.

Planning also involves reviewing the broader software landscape.
Conduct a survey of existing tools, frameworks, or libraries to avoid reinventing the wheel.
Integrating well-tested third-party components can save time and ensure better functionality.
Regularly revisit and refine the plan as the project progresses, keeping it aligned with the overarching research goals.

%\textbf{Legacy Projects}. 
\update{Planning is also important for legacy projects, but the problem statements may be different. What are the pain points of this project (e.g., compile time, testing speed)? How should the team prioritize their backlog of open tasks? It can be especially helpful to explore previous choices made and reevaluate which ones are still valid or should be changed.}

\update{As an example, one of the authors recently undertook a large effort to redesign a fundamental interface in a legacy software package. The author met with senior developers over several months to target specific questions such as, ``What are the current maintenance challenges? What do we need to incorporate from the current interface? What should be deprecated or removed?'' These discussions led to a list of desired changes, priorities, and requirements that could be transformed into a project plan.}

\subsection{\textbf{\textit{Guideline 2: Design for Modularity}}}
\label{rule-modularity}

Modular design is about building software as a collection of independent, reusable components.
Start by identifying the core functionalities of your software and structuring them into self-contained modules.
%\strike{For example, separate data handling, computation, and visualization into distinct parts, each with a clear interface.}
\update{For example, you might have one module for handling data, another for performing calculations, and a third for displaying results, with each module focusing on a specific function.}
% I think this language is simpler / more tractable for the average reader.

Encapsulation is key to modularity, \update{where each module has clear inputs (what it needs to work) and outputs (what it produces), sharing only the necessary information for interaction}.
%\strike{Each module should have well-defined inputs and outputs, exposing only what is necessary for interaction.}
% Again, this is slightly simpler language that is more accessible.
This reduces dependencies between modules, making it easier to debug and update individual components.
For instance, if a visualization module needs an update, you should not need to alter the data processing logic.

Think about reusability beyond your project.
Writing generic, well-documented, and well-tested modules can allow them to be reused in other projects, extending their impact.
Use dependency injection \update{-- a technique where a module receives the tools it needs from outside rather than creating them itself --} where appropriate to make modules more flexible, allowing them to work with different implementations without modification.

%\textbf{Legacy Projects}. 
\update{Creating modularity in legacy projects where it does not already exist is no easy task -- but that does not mean it is impossible. Begin by assessing and identifying the different overall functionalities of the code and exploring how they connect. Then, make a plan (see \nameref{rule-plan}) on how these elements can be separated. Focus on finding opportunities to combine like-features into the same or related modules (e.g., move all functionality related to file system interaction into one core location). Write tests and document as you go, keeping a trail so anything can be undone if the functionality breaks. It won't be a fast process, so approach it gradually and modify only one type of functionality at a time.}

\subsection{\textbf{\textit{Guideline 3: Write Clean and Readable Code}}}
\label{rule-cleancode}

Clean code is not only for aesthetic purposes~\cite{martin2009clean}; it directly impacts the readability, maintainability, and longevity of the software. Start by adopting a consistent coding style and stick to it throughout the project. Consistency makes it easier for new collaborators to understand and contribute to the codebase. Use tools like linters\footnote{\url{https://www.sonarsource.com/learn/linter/}} (e.g., Flake8, ESLint, clang-tidy, or fortitude) to automatically enforce coding standards and identify issues early.
\update{Coding standards are agreed-upon guidelines that define how code should be written and formatted to ensure consistency, readability, and maintainability across a project or team.} 

Avoid over-complicating the code. Strive for simplicity by breaking down tasks into small, manageable functions or methods. Each function should do one thing well. Functions and variables should have descriptive names that reflect their purpose. \update{For example, avoid using generic variable names like ``i'' for loop iterators, which provide little context of its intent.} If you find yourself writing the same or similar code in multiple places, refactor it into reusable components to reduce duplication \update{(see \nameref{rule-modularity})}.

Readable code minimizes reliance on excessive comments, but comments still have their place. Use them to explain the ``why'' rather than the ``what'' when the purpose of the code might not be immediately clear. Peer code review (see~\nameref{rule-codereview}) can help spot code that is hard for others to understand. Additionally, invest in learning and applying design patterns where appropriate. \update{Design patterns are reusable blueprints for common software design challenges.} 
%\strike{Patterns such as Model-View-Controller (MVC) or Factory Method can add structure and clarity to the code, especially in larger projects.}

%\textbf{Legacy Projects}. 
\update{When considering legacy projects, modifying the code to make it clean and readable can be a major step. Legacy code is often hard to understand due to many factors such as lack of standard style, generic or even reused variable names, having large complex functions, and dead code that is no longer used.}

\update{Agree on a standard style for the codebase. Start by using this standard for any new code added to the project. If you modify existing code to use the standard, avoid making these changes simultaneously with changes to any functionality improvements.}

\update{Rename variables to be more descriptive.
Remove code that is no longer used. Write comments where needed. Keep functions short and focused. Avoid duplications. These are some of the simplest things you can do to improve readability.}

\subsection{\textbf{\textit{Guideline 4: Use Version Control}}}
\label{rule-versioncontrol}

Version control is not just about tracking changes. It is about fostering collaboration and maintaining a history of the software’s evolution. Start by organizing your repository logically, separating core code, test scripts, data, and documentation into clear directories. This structure helps maintain clarity as the project grows.

When collaborating, agree on a branching strategy upfront, such as Git Flow or trunk-based development. These strategies define workflows for feature development, bug fixes, and releases, reducing confusion among contributors. Use pull requests for code reviews, ensuring that all changes are vetted before merging them into the main branch. This practice improves code quality and provides a learning opportunity through peer feedback (see \nameref{rule-codereview}).  Pull requests are also often used in conjunction with continuous integration testing (see \nameref{rule-test}).

For open-source projects, maintaining a clear record of contributions and contributors is vital. Practice clearly documenting how potential contributors should engage with the project via the version control repository, including cloning the repository, setting up their environment, and proposing changes. This kind of documentation often appears in the project's README file at the top level of the repository or in a separate file with a name like CONTRIBUTING. Additionally, create a changelog to document updates in each version, ensuring transparency and helping users understand how the software evolves.\footnote{Not sure how to get started? Visit \url{https://intersect-training.org/collaborative-git/} for a good beginner's course.}

%\textbf{Legacy Projects}. 
\update{If legacy code is not already within a Git repository, consider migrating it to a Git platform. While older version control systems (VCS) such as Subversion and CVS are adequate to keep track of code changes, they lack the collaborative features (e.g., pull requests) that Git and platforms such as GitHub, GitLab, and Bitbucket offer. These modern platforms effectively support collaborative, geographically distributed development.}

\update{One of the authors has successfully migrated multiple legacy codes from Subversion using the \emph{git-svn} tool, which, in turn, increased collaboration in software development. Two elements are critical to successfully migrating an existing code base from another VCS to Git:}
\update{\begin{enumerate}%
   \item \textbf{Agree on a simple but straightforward workflow}: Establish a clear and consistent workflow that all team members can follow.%
   \item \textbf{Provide comprehensive support}: Provide clear documentation detailing Git commands and workflows. Additionally, provide individual support to developers new to Git, as adapting from a centralized Subversion workflow to a distributed Git workflow takes time.%
\end{enumerate}}

\subsection{\textbf{\textit{Guideline 5: Test Your Code Regularly}}}
\label{rule-test}

Testing is an integral part of the software development lifecycle, not an afterthought. Start with unit tests for individual components, ensuring each works as intended. Complement these with integration tests that verify the interoperability of different modules. For large-scale software, end-to-end tests simulate real-world usage scenarios, validating the software from input to output.

Automated testing frameworks are invaluable for maintaining quality as the codebase grows. Continuous integration (CI) \update{is defined as the practice of integrating code into the codebase while ensuring a stable codebase.} CI tools like GitHub Actions or Jenkins can automatically run your tests whenever code is pushed, catching errors early. CI testing is often used to vet pull requests before they are merged. Beyond functionality, consider performance tests to ensure the software meets speed and efficiency requirements, especially for computationally intensive tasks. When relying on third-party libraries or modules, consider writing tests for key functionality that your code depends upon.  Such tests can be used to verify that it is safe to update the dependency, for example.

Documentation of the testing process is equally important. Include instructions for running tests and interpreting their results in your repository. Also, consider using code coverage tools to identify parts of the code that lack test cases, ensuring comprehensive testing.

The practice of writing tests is a strategic investment of time. Although it may initially appear to slow progress, this effort will yield substantial benefits in the long term, enhancing both the reliability and maintainability of the codebase. Moreover, it can significantly reduce, or even eliminate, the need for extensive debugging. Consider reevaluating your testing practices at strategic points in your project's growth to right-size your practices to your current project maturity \cite{mundt22issc}.

%\textbf{Legacy Projects}
\update{While legacy code often lacks an adequate test suite, it is rare for it not to have at least some examples. You can use these examples to build integration tests that allow you to compare the outputs and performance of those examples before and after changes to the code. Your first task may be to gather these examples from different researchers. Working with field experts to design these tests is essential, as they can often point to analytical solutions or published test cases.}

\update{If you need to familiarize yourself with the existing codebase, strategically writing new unit or integration tests that exercise selected parts of the code is a great interactive way to learn the code. Good test coverage is critical when extending code you don't know well because they reduce the anxiety of breaking the code. One easy way to expand your test suite is by conducting compatibility or cross-platform tests. For instance, compile your code with different compilers (e.g., GNU, Intel), run it with different Python versions, different operating systems, or on different hardware. Often, different environments will catch different bugs.}

\update{Once you have some tests, you need to run them automatically. One common mistake is only testing what was changed, which might overlook some side effects. Depending on the computing resources the tests require, you may need to run different sets of tests on different schedules. For instance, a short fifteen-minute test can run for each commit, a medium one-hour test can run daily, and a more extended four-hour test can run weekly. If a change breaks the code, it's easier to trace back through the developments introduced last week.}

\update{Finally, if your task is to add new code, agree on modern testing policies for new code with unit tests.}

\subsection{\textbf{\textit{Guideline 6: Make Peer Code Review a Standard Practice}}}
\label{rule-codereview}

Peer code review is a cornerstone of producing high-quality research software. It involves having colleagues or team members examine code changes to identify errors, ensure compliance with coding standards, and improve overall quality. This practice not only helps catch bugs and inefficiencies early but also ensures that the code is maintainable, changes align with the project's goals, and knowledge of functionality is shared between colleagues. In research contexts, where the accuracy and reproducibility of results depend on the integrity of the software, peer code review serves as an essential safeguard. Bringing multiple perspectives to the review process reduces the risk of oversight and improves the robustness of the codebase.

The benefits of peer code review extend beyond bug detection. It creates an environment where team members learn from one another, sharing knowledge about coding practices, domain-specific logic, and design patterns. This collaborative process fosters a culture of accountability and continuous improvement, where developers strive to write cleaner, more efficient code, knowing it will be reviewed. For junior team members, code review is an invaluable learning tool, exposing them to better practices and giving them a safe space to receive constructive feedback. For senior team members, it offers the opportunity to mentor others and learn about new techniques or software engineering methods. Code reviews also ensure that multiple team members understand all parts of the codebase. This practice mitigates the risk associated with the departure of any single developer, thereby reducing the potential impact on the project's continuity and stability.

To implement effective peer code review, establish a structured workflow. Platforms like GitHub, GitLab, or Bitbucket make this process straightforward, allowing developers to submit pull requests for changes and reviewers to comment directly on the code. Set clear criteria for reviews, focusing on functionality, readability and understandability, adherence to coding standards, test coverage, and corresponding documentation updates. To streamline code reviews and reduce review time, focus on specific issues or tasks rather than reviewing large code chunks. Automate routine checks, such as style and syntax compliance, using tools like linters, so reviewers can focus on higher-level concerns. Cultivate a healthy code review environment by avoiding assertive language and emphasizing constructive feedback, where reviewers not only point out issues but also suggest improvements and provide positive affirmations. By embedding peer code review into the development cycle, research teams can elevate the quality and reliability of their software, ensuring it meets both scientific and technical excellence.\footnote{Not sure how to get started? Visit \url{https://intersect-training.org/Code-Review/} for a good beginner's course.}

%\textbf{Legacy Projects.}}
\update{Legacy codebases often pose unique challenges, such as lack of documentation, outdated coding styles, or inconsistent practices, but peer code review can play a transformative role in gradually improving their quality. Introducing peer review into a legacy project does not require a full rewrite or even a complete review of the existing code. Instead, you can adopt an incremental approach, reviewing changes as they are made during bug fixes, feature additions, or refactoring efforts. This ``Boy Scout Rule'' approach \textit{always leaves the code better than you found it} helps modernize the codebase over time without halting development.}

\update{Reviewing legacy code also helps surface hidden assumptions, undocumented features, and potential risks. You can gain a better understanding of how older components work, which improves collective code ownership and knowledge sharing. Over time, this process can standardize practices across the codebase, making it easier to maintain and extend. Moreover, applying peer review to legacy code changes ensures that improvements align with current standards, helping to avoid perpetuating technical debt. By embedding peer review into the workflow for legacy systems, teams can enhance code reliability, reduce long-term maintenance costs, and improve the overall sustainability of the software.
}

\subsection{\textbf{\textit{Guideline 7: Document Everything}}}
\label{rule-document}

Documentation is often overlooked in research software, yet it is critical for usability and reproducibility. Begin with a README file that provides an overview of the software, its purpose, and how to get started. Include installation instructions, supported platforms, dependencies, and prerequisites to set user expectations.

More generally, think about different kinds of documentation that support both developers and users of the software -- even if you are the only developer \emph{or} user, your ``future self'' will thank you. 
Document your development process, coding standards, and \update{any workflows in place. Workflows are the processes used to complete a specific software-related goal. They can include steps to perform data analysis, the code review process, how to contribute to the code, etc.} Capturing the requirements that have been defined and architectural decisions that have been made can be invaluable in keeping developers on the intended track.  Tools like Doxygen or Sphinx allow some forms of developer documentation embedded in the code to be extracted in a more reader-friendly form.  This can make it easier for developers to keep the documentation current as the code evolves (but diligence is still needed -- see \nameref{rule-codereview}).

For users, consider both the user guide and tutorial materials. The user guide should detail the features and capabilities and how to invoke them, preferably with examples, as well as troubleshooting tips and frequently asked questions.  Tutorials, on the other hand, walk the reader through a series of steps to accomplish specific goals.  Tutorials can be produced in many different formats -- text, presentation slides, or even videos.

Interactive documentation, such as Jupyter notebooks with embedded code examples, can be particularly effective for research software. These allow users to experiment with the software and understand its functionality in a hands-on manner.

%\update{\textbf{Legacy Projects.}}
\update{Legacy projects often suffer from poor or missing documentation, making them difficult to maintain, extend, or even use. Introducing systematic documentation practices into a legacy codebase is a practical and low-risk way to improve its quality and longevity. Start by documenting what is, not just what should be. Reverse engineer the system by recording how the current software works: key modules, data flows, input/output formats, and known limitations. Even brief summaries or high-level diagrams can make a significant difference in onboarding new developers or revisiting the project after a long break.}

\update{Incrementally update or create documentation as you work on specific parts of the legacy code. For example, when fixing a bug or adding a feature, add comments or docstrings to the functions you touch, update the README sections, or clarify usage examples. This progressive enhancement aligns well with the Boy Scout Rule — making the system slightly better with every interaction.}

\update{Legacy systems also benefit greatly from decision documentation: recording why certain approaches or libraries were used, what trade-offs were made, or which alternatives were considered. This historical context helps avoid repeated mistakes and ensures future changes respect past constraints. You can capture this in docs/architecture.md, changelogs, or internal wikis.}

\update{For users, improving documentation in legacy systems can make them more accessible and usable. Even if the software was not originally designed with usability in mind, adding clear guides, command-line usage examples, or simple Jupyter notebooks can open the project to a broader audience, including collaborators and non-specialist users. Tutorials and FAQs help reduce support burdens and make the project more sustainable in the long term.}

\subsection{\textbf{\textit{Guideline 8: Strive for Reproducibility}}}
\label{rule-reproducibility}

Reproducibility is essential for ensuring the credibility and impact of research in general and research software in particular~\cite{reproducibility_in_science}, \update{and it has links to a number of the other rules in this paper}. Start by defining a clear workflow for setting up the software environment. Tools for containerization like Docker or virtualization like Venv or Conda can help encapsulate dependencies, ensuring consistency across systems.

Include detailed instructions for reproducing results, from data preprocessing to final analysis. For complex data analysis workflows, use workflow management tools like Snakemake or Nextflow to automate the process. This not only ensures reproducibility but also saves time for both the developer and the user.

\update{The configuration files for your containers or virtual environments, as well as the scripts for your workflows, can serve as a form of documentation of your ``experimental'' setup and processes (see \nameref{rule-document}).  They should be kept under version control (see \nameref{rule-versioncontrol}). Also, it is important to use only well-defined versions of any code in an experiment (i.e., no local changes from what is in the repository) and document the versions used for each experiment.}

For datasets, provide versioned archives or use data repositories like Zenodo. If the data cannot be shared due to privacy or legal constraints, provide synthetic datasets or detailed descriptions of how the data was prepared. Maintaining a clear, reproducible workflow fosters trust in your results and allows others to build upon your work.

If your software uses concurrency, such as multiple threads on an individual node or multiple processes distributed across multiple nodes communicating through MPI, be aware of the effects that the non-deterministic order of mathematical operations can have on the results, and what reasonable criteria are for their reproduction from one run to another or one platform to another.

\update{It is also worth continuing to run your test suite periodically on the experimental platform (see \nameref{rule-test}) to reconfirm that it is working as expected. Software updates to the platform (e.g., operating system, lower-level libraries, and tools you depend on) can introduce issues and impact the reproducibility of your work, especially for longer-running experimental campaigns.}

\update{If you have inherited a code or are new to a project, attempting to reproduce some prior results may be a useful intermediate-level exercise to gain familiarity and comfort with the code (after successfully building the code and running through the available examples or tutorials).  It will also allow you to test and refine the team's reproducibility-oriented practices for the future.}

\update{An extensive set of suggestions for improving reproducibility at all stages of the process of computationally-based research can be found in the ``Improving Reproducibility through Better Software Practices'' module\footnote{See \url{https://doi.org/10.6084/m9.figshare.26384188}, file \texttt{08-reproducibility.pdf}} of the Better Scientific Software tutorials.\footnote{\url{https://bssw-tutorial.github.io/}}}

% \subsection{\textbf{\textit{Rule 8: Design for Modularity}}}
% \label{rule-modularity}

% Modular design is about building software as a collection of independent, reusable components. Start by identifying the core functionalities of your software and structuring them into self-contained modules. For example, separate data handling, computation, and visualization into distinct parts, each with a clear interface.

% Encapsulation is key to modularity. Each module should have well-defined inputs and outputs, exposing only what is necessary for interaction. This reduces dependencies between modules, making it easier to debug and update individual components. For instance, if a visualization module needs an update, you should not need to alter the data processing logic.

% Think about reusability beyond your project. Writing generic, well-documented modules can allow them to be reused in other projects, extending their impact. Use dependency injection where appropriate to make modules more flexible, allowing them to work with different implementations without modification.

\subsection{\textbf{\textit{Guideline 9: Pay Attention to Performance and Scalability}, as Necessary}}
\label{rule-performance}

As research problems grow in complexity, the performance and scalability of the software become critical. During the development cycle, consider the computational requirements of your target problems. Ensure your design and algorithms are efficient within your resource constraints.  If it looks problematic, you may want to look for other approaches or different algorithms that may be faster or require fewer resources.  If that is not possible, you may need to use on-node concurrency (e.g., threading on CPUs or GPUs), or expand to multiple nodes using, for example, MPI to exchange data between processes. Frameworks like Dask (Python) or Spark can handle large datasets more efficiently than traditional single-threaded approaches.

Begin by profiling the code to identify slow or resource-intensive sections. Tools like cProfile (Python), Perf (Linux), or VisualVM (Java) can help pinpoint bottlenecks. Once identified, optimize these areas by replacing inefficient algorithms or reducing unnecessary computations.  Remember, however, that it is not productive to spend time optimizing portions of the code that are not a bottleneck for the problems you are trying to solve -- speeding up a portion of the code by a factor of 10 has little effect if that code constitutes less than one percent of the overall runtime.

For large-scale computations, consider using parallel processing, GPU acceleration, or distributed systems.  Always measure the performance impact of optimizations to ensure they achieve the desired results without introducing new issues.

Scalability involves designing the software to handle increasing workloads gracefully. For example, modular architecture can make it easier to scale individual components. Additionally, document performance benchmarks so users know the limits of the software and can plan their use accordingly.

At the same time, it is essential to remember the well-known quip from computer scientist Tony Hoare, who famously remarked, ``Premature optimization is the root of all evil''~\cite{hoare1981emperor}. This underscores the importance of focusing first on writing clean, clear, and understandable code, accompanied by comprehensive tests. Optimization, which often reduces code readability and maintainability, should only be pursued when necessary and with careful consideration. When optimizing, use your tests to verify that the modified code continues to produce correct and reliable results.

\update{Performance and scalability are typically important throughout the lifecycle of research software, so legacy software that may have been inherited from others is subject to the same considerations.  However, in this context, optimizations already done in the code may have made it harder to read and understand. It may be worth investigating the version control history and/or the documentation to find more straightforward, more easily understood versions of algorithms, or at least documentation pertaining to the optimization to aid in understanding it better. If the usage of the code has changed so that previously optimized code is no longer a performance bottleneck, developers might even consider reverting to a more straightforward implementation for the sake of maintainability (see \nameref{rule-maintenance}).}

\update{Many computer centers such as National Energy Research Scientific Computing Center (NERSC), Oak Ridge Leadership Computing Facility (OLCF), Texas Advanced Computing Center (TACC), San Diego Supercomputer Center (SDSC), Leibniz Supercomputing Centre (LRZ), and Swiss National Supercomputing Centre (CSCS) offer training, hands-on assistance, and other resources for performance engineering and parallelization of research software to help get started effectively.}

\subsection{\textbf{\textit{Guideline 10: \update{Plan for Long-Term Maintenance}}}}
\label{rule-maintenance}

Software maintenance is as important as its initial development, ensuring continued usability and relevance. 
\update{Even software that is initially thought to be ``throw away'' often lives much longer than anticipated.} It is, therefore, useful to make the common maintenance tasks as easy as possible to carry out. \update{The previous guidelines will contribute to a well-maintained code. Specifically,  automating repetitive tasks such as testing and deployment using Continuous Integration and Continuous Delivery (CI/CD) pipelines \update{(see \nameref{rule-test})} reduces the burden of manual updates and catches problems early} Here, Continuous Delivery (CD) is the practice of automating the deployment of tested code into a test or production-level environment.

Keep the software dependencies up-to-date, as outdated libraries can lead to security vulnerabilities or compatibility issues. Tools like Dependabot or pip-tools can simplify this process. Additionally, document the development process, including coding standards, all workflows, and system architecture, to make it easier for new contributors or maintainers to take over \update{(see \nameref{rule-document})}.

For open-source projects, ensure that the community has clear guidelines for contributing and that the codebase is well-documented. \update{Develop a CONTRIBUTING.md file that outlines the process for contributing to the project, including coding standards, branching strategies, and how to submit pull requests. Encourage community contributions by labeling issues that are good for newcomers. Host regular community meetings to discuss project progress and gather feedback.} If you can no longer maintain the project, consider archiving it with clear instructions for reuse or forking, ensuring its value persists even after active development ends.

% \textbf{Legacy projects}.
\update{Legacy projects can be difficult to maintain. Without good tests, it can be hard to make modifications with confidence. If code and workflows are not well documented, it can be overwhelming for new contributors. Inconsistent coding styles and large complex functions can make things difficult to read and understand. Adherence to the guidelines mentioned here and previously can be helpful for addressing these difficulties.}

\update{Consider taking your existing code and building some basic CI/CD pipelines. Think about how your code is currently configured, built, and installed, then automate this process. This will help verify your software continues to work during any changes you decide to implement, based on the guidelines within this document.}

%% file: sec_conclusion.tex
High-quality research software is an essential component of modern scientific inquiry, enabling the analysis of complex data, reproducibility of results, and the development of innovative solutions. This paper presented ten essential guidelines for producing research software that is robust, maintainable, and impactful. 
By following these guidelines, researchers can overcome common challenges such as inadequate usability and short-lived software lifespans. High-quality research software not only supports immediate scientific goals but also contributes to a broader ecosystem of tools that facilitate collaboration and accelerate scientific discovery. As science becomes increasingly data-driven and interdisciplinary, producing reliable and reusable software is not merely a technical achievement but a vital contribution to the progress of science. Researchers who adopt these principles will maximize the impact of their software and play a pivotal role in advancing a culture of excellence in scientific computing.

%% file: acknowledgments.tex
This material is based in part upon work supported by the U.S. Department of Energy, Office of Science, Office of Advanced Scientific Computing Research, Next-Generation Scientific Software Technologies program, under contract number DE-AC05-00OR22725 to ORNL.

Sandia National Laboratories is a multimission laboratory managed and operated by National Technology \& Engineering Solutions of Sandia, LLC, a wholly owned subsidiary of Honeywell International Inc., for the U.S. Department of Energy’s National Nuclear Security Administration under contract DE-NA0003525.

%% file: author-bios.tex
\begin{IEEEbiography}{Nasir U.~Eisty} is an Assistant Professor of Computer Science at the University of Tennessee, Knoxville. His research interests lie in the areas of Software Engineering, AI for Software Engineering, Research Software Engineering, and Software Security. Eisty received his Ph.D degree in Computer Science from the University of Alabama. Contact him at neisty@utk.edu
\end{IEEEbiography}

\begin{IEEEbiography}{David E.~Bernholdt} is a distinguished R\&D staff member at Oak Ridge National Laboratory.  His research interests are broadly in the development of scientific software for high-performance computers, including developer productivity, and software quality and sustainability. Bernholdt received his Ph.D. degree in chemistry from the University of Florida. Contact him at bernholdtde@ornl.gov.
\end{IEEEbiography}

\begin{IEEEbiography}{Alex Koufos (he/him)} is a skilled Research Software Engineer (RSE) and computational physicist with a diverse background in quantum mechanics, software engineering, and data science. His career has spanned various industries, including software engineering, where he contributed to the development of 3D visualization software for training purposes, and autonomous robotics. Currently, Alex is dedicated to maintaining the data infrastructure for the Solar Dynamics Observatory (SDO) mission at Stanford University. He holds a Ph.D. in Computational Quantum Mechanics from George Mason University, where he focused on superconductivity research through first-principles simulations. Contact him at akoufos@sun.stanford.edu.
\end{IEEEbiography}

\begin{IEEEbiography}{David J. Luet} is an Associate Director of the Research Software Engineering group at Princeton University, where he leads a team developing sustainable and reproducible research software. His work focuses on advancing Princeton's research and educational missions through innovative methodologies and strategic initiatives. Luet received his Ph.D. in Mechanical and Aerospace Engineering from Princeton University. Contact him at luet@princeton.edu.

\end{IEEEbiography}

\begin{IEEEbiography}{Miranda Mundt (she/her)} is a Research Software Engineer (RSE) and senior member of technical staff with the Department of Software Engineering and Research, Sandia National Laboratories. As a practitioner, researcher, and consultant, she works closely with scientific software teams across various domains including machine learning benchmark development and optimization modeling to advance software engineering in scientific computing. She is a certified Carpentries instructor and has led multiple tutorials on better software practices including documentation, reproducibility, and traceability. Mundt received her B.S. degree in applied mathematics from the University of New Mexico. Contact her at mmundt@sandia.gov.
    
\end{IEEEbiography}